\documentclass[twocolumn,superscriptaddress,aps,showpacs,floatfix,prl,10pt]{revtex4-1}
\usepackage[T1]{fontenc}
\usepackage{amssymb,amsmath,bm}
\usepackage{graphicx}
\usepackage{color}
\usepackage{bbold}
\usepackage{bbm}
\usepackage{appendix}
\usepackage{soul}
\usepackage[dvipsnames]{xcolor}
\usepackage{ulem}
\usepackage{float}
\usepackage{textcomp}
\usepackage{ mathrsfs }
\usepackage{siunitx}%

\newcommand{\RNum}[1]{\uppercase\expandafter{\romannumeral #1\relax}}

\begin{document}
\title{Non-reciprocal Exciton-Polariton Ring Lattices}

\author{Huawen Xu}
\email{huawen001@e.ntu.edu.sg}
\affiliation{Division of Physics and Applied Physics, School of Physical and Mathematical Sciences, Nanyang Technological University, 21 Nanyang Link, Singapore 637371, Singapore}
\author{Kevin Dini}
\affiliation{Division of Physics and Applied Physics, School of Physical and Mathematical Sciences, Nanyang Technological University, 21 Nanyang Link, Singapore 637371, Singapore}
\author{Xingran Xu}
\affiliation{Division of Physics and Applied Physics, School of Physical and Mathematical Sciences, Nanyang Technological University, 21 Nanyang Link, Singapore 637371, Singapore}
\author{Rimi Banerjee}
\affiliation{Division of Physics and Applied Physics, School of Physical and Mathematical Sciences, Nanyang Technological University, 21 Nanyang Link, Singapore 637371, Singapore}
\author{Subhaskar Mandal}
\affiliation{Division of Physics and Applied Physics, School of Physical and Mathematical Sciences, Nanyang Technological University, 21 Nanyang Link, Singapore 637371, Singapore}
\author{Timothy C. H. Liew}
\email{timothyliew@ntu.edu.sg}
\affiliation{Division of Physics and Applied Physics, School of Physical and Mathematical Sciences, Nanyang Technological University, 21 Nanyang Link, Singapore 637371, Singapore}
\affiliation{MajuLab, International Joint Research Unit UMI 3654, CNRS, Universit\'e C\^ote d'Azur, Sorbonne Universit\'e, National University of Singapore, Nanyang Technological University, Singapore}

\begin{abstract}
Recent experiments have shown the transfer of orbital angular momentum (OAM) from a non-resonant laser onto an exciton-polariton condensate, despite earlier views that the phase information of such a laser should be lost during the process of polariton condensation. We study with a phenomenological theory the interplay of a usual angular momentum independent gain and an angular momentum preserving gain. We find that even when the latter is much smaller, it is enough to favour condensation into a given orbital angular momentum state. This further allows a breaking of symmetry in the system, which further manifests in non-reciprocal one-way propagation in a lattice of coupled rings. Even though we consider only Hermitian reciprocal coupling between rings, the local non-Hermiticity generates an effective non-reciprocal coupling and supports a non-Hermitian topological invariant (winding number) associated to a non-Hermitian skin effect.

\end{abstract}
\maketitle

\emph{Introduction}.
Exciton-polariton condensation is a process characterized by the spontaneous formation of coherence and the spontaneous breaking of $U(1)$ phase symmetry. Experiments demonstrating this effect have typically used a non-resonant laser to excite first electron-hole pairs above the bandgap, which can relax in energy to form a polariton condensate~\cite{Byrnes2014,Carusotto2013,Deng2002,Deng2010,Kasprzak2006}. Even though the laser itself already has coherence, it has been widely accepted that the large sequences of scattering processes involved in the relaxation of electron-hole pairs would lose such coherence, such that a spontaneous reformation of coherence and spontaneous choice of phase remains a necessary feature of polariton condensation. This is consistent with the observation that polariton condensation requires a threshold density, below which polaritons are incoherent despite the presence of the coherent laser.

Given that the phase coherence of the laser is lost during polariton condensation, one could na\"{i}vely think that the polarization of the laser is also lost; after all a polarization corresponds to a definite phase relation between orthogonal components. However, experiments~\cite{Ohadi2012,Kammann2012} showed that it is possible to transfer the circular polarization of the non-resonant laser onto the circular (spin) polarization of a polariton condensate. The logical explanation is that the optically oriented electron-hole pairs excited by the laser, while losing phase coherence, may preserve their spins. Even though the spin polarization of an electron-hole pair reservoir is expected to be incomplete (e.g., in~\cite{Carlon-Zambon2019} an 18$\%$ spin polarization was measured in bare quantum well structures), polariton condensates may achieve 100$\%$ circular polarization~\cite{Klaas2019}. In contrast, no works have claimed the transfer of the linear polarization of a non-resonant laser onto a polariton condensate. The transfer of spin polarization under non-resonant laser excitation was described with a phenomenological mean-field model~\cite{Kammann2012}, and similar models with the same form of spin conserving gain fitted well with later experiments~\cite{Anton2015,Cilibrizzi2015,Askitopoulos2016}. 

In analogy to the transfer of spin, one can also ask whether it is possible to transfer the OAM of a non-resonant laser onto a polariton condensate. Again, the na\"{i}ve answer would be negative, as the OAM of an optical field corresponds to an angular phase gradient, implying phase coherence. However, a recent experiment shows that the opposite occurs~\cite{Kwon2019}: the OAM of a non-resonant laser can be transferred to a polariton condensate. This is in contrast to the generally accepted mean-field theory of polariton condensation~\cite{Wouters2007,Keeling2008,Carusotto2013}, where the effect of the non-resonant laser is treated as a phenomenological gain proportional to its intensity. The intensity may vary with the position in the microcavity plane, but, such an intensity profile alone can not define an angular momentum if it is circularly symmetric (which is the case for typical OAM carrying laser beams). Aside the discrepancy with existing theory, the ability to maintain the OAM of a non-resonant laser is promising for the generation of vortex polariton states, which were previously thought to require a chiral arrangement of multiple spots~\cite{Dall2014}. 

\begin{figure}[t]
\includegraphics[width=1\columnwidth]{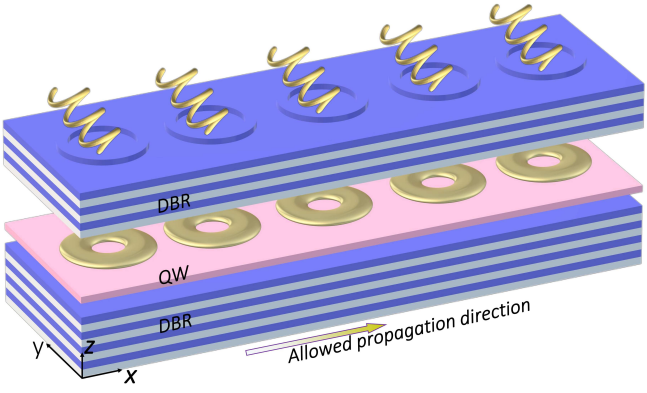}
\caption{Schematic of a microcavity with edged ring lattices under non-resonant OAM preserving pump and the exciton-polariton vortex condensation is formed. A quantum well (QW) is placed in the middle of two sets of distributed Bragg reflectors (DBRs). Due to the non-reciprocal coupling between separate rings, the propagation is only allowed in one direction.}
\label{Figure1}
\end{figure}

Instead of a planar microcavity, if polaritons in a ring geometry are considered then the transfer of  the OAM breaks the reciprocity as the ring favours a mode of given circulation (clockwise or counter-clockwise).  Such a  non-reciprocal ring is known as the circulator in classical wave systems~\cite{Fleury2014,Kord2020}, which relies on an effective magnetic field and is extremely useful in many practical applications, such as input-output isolation. Here, we consider a chain of such rings (see Fig.~\ref{Figure1}), which could be formed from the technique of etching~\cite{Dreismann2014,Mukherjee2019,Wang2021}. Transferring the non-reciprocity from the rings to the whole chain is non-trivial as any ring would couple to its neighbours in both directions. Nevertheless, we will show that through placement of defects around the rings at specific angles, which couple the clockwise and anticlockwise modes with a  specific phase, the chain becomes non-reciprocal. Numerical modelling shows that non-resonantly injected polaritons propagate only in one way through the chain, while the propagation in the opposite direction is restricted. We further show that the obtained one way propagation is  associated with the topological non-Hermitian skin effect~\cite{Gong2018,Lee2019,Zhang2020,Zhu2020,Zhang2021,Mandal2020,Xu2021,Banerjee2021,Mandal2021}, which can be characterized by the behaviour where all eigenmodes of the system localize at one edge and carries an associated topological invariant. The one-way propagation is highly relevant for the development of computational schemes based on polariton angular momentum modes~\cite{Sigurdsson2014,Ma2017,Gao2018,Ma2020}.

\emph{OAM Preserving Non-Resonant Excitation.}
To model a non-resonant pump carrying OAM, we assume two components: a component that does not preserve OAM and can be treated with the usual Gross-Pitaevskii (GP) equation coupled to an exciton reservoir; an additional term corresponding to a gain of an OAM carrying mode with profile equivalent to that of the non-resonant pump. The evolution of the polariton wavefunction $\psi(\bm{r},t)$ and the exciton reservoir density $n_R(\bm{r},t)$ is given by:
\begin{align}
i\hbar \frac{\partial \psi(\bm{r},t)}{\partial t}=\bigg(-\frac{\hbar^2 \nabla^2}{2m} 
 +g |\psi(\bm{r},t)|^2 + (g_{R}+\frac{i\hbar R}{2})n_{R}(\bm{r},t) 
\notag \\
  -i\gamma \bigg) \psi(\bm{r},t) + iF_0\iint \phi^{*}(\bm{r'}) \psi(\bm{r'},t)d\bm{r}' \phi(\bm{r}),  \notag \\
\frac{\partial n_R(\bm{r},t)}{\partial t}= (-\gamma_R + R|\psi(\bm{r},t)|^2)n_R + P(\bm{r}).
\label{Equation1}
\end{align}
where $m$ is the polariton effective mass, $\gamma$ is the polariton decay rate, $g$ and $g_R$ describe the strength of polariton-polariton and polariton-exciton interactions, $R$ is the condensation rate, and $\gamma_R$ is the reservoir decay rate. Apart from the integral term, the above coupled equations have been used to successfully demonstrate the physics related to vortices in exciton-polariton system~\cite{Ma2017,Ma2017Soliton,Ma2018}. The OAM carrying pump term, corresponding to the integral in Eq.~\ref{Equation1} can be derived by decomposing the wavefunction in a basis of modes containing $\phi(\mathbf{r})$, which is a mode desired to be pumped, and then transforming back to the original basis in real space (see supplementary material [SM], Ref.~\cite{SupportingInformation}). The mode to be pumped is chosen as $\phi(\mathbf{r})=e^{-i\theta}r^2e^{-(\mathbf{r}-\mathbf{r}_0)^2/\delta r^2}$, where $\mathbf{r}_0$ is the ring center, $\theta$ an angular coordinate around the ring center, and $\delta_r$ determines the ring size. The pumping of the exciton reservoir population should be a real number and so is chosen as $P(\mathbf{r})=P_0|\phi(\mathbf{r})|^2$.  

\begin{figure}[h!]
\includegraphics[width=1\columnwidth]{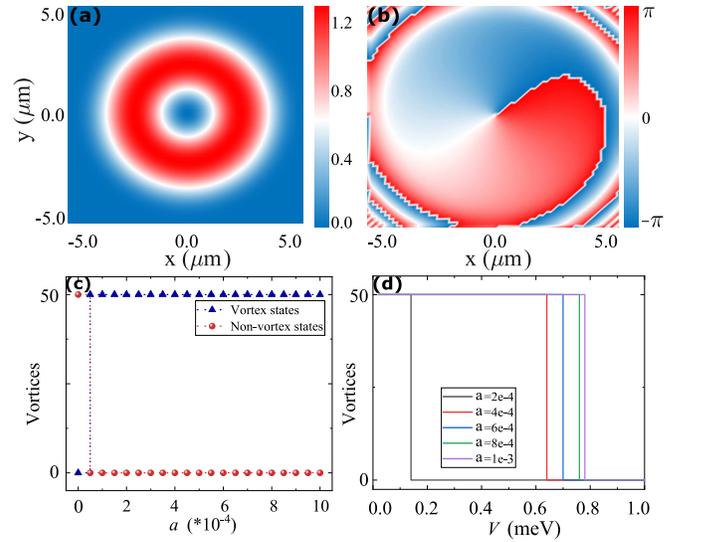}
\caption{\textbf{a}) Intensity of the polariton wavefunction $|\psi(\bm{r})|^2$ in the steady state (colour scale in $\mu m^{-2}$), $a$=$10^{-4}$. \textbf{b}) Phase distribution of the vortex state, $a$=$10^{-4}$. \textbf{c}) Number of vortex/non-vortex states among 50 random realization under different ratio $a$. \textbf{d}) Number of vortex states among 50 random realization under different disorder root mean square amplitude ($V$). Parameters: $g=1$ $\mu eV \mu m^2$, $g_R = 2g$, $\gamma=0.2$ $meV$, $\gamma_C = 5\gamma$, $R=0.05$ $ps^{-1}\mu m^2$, and $m=5\times10^{-5}m_e$, with $m_e$ the free electron mass.}
\label{Figure2}
\end{figure}

We will assume that the ratio of the OAM conserving and non-conserving pump $a=F_0/P_0 \hbar$ is small. This is consistent with the fact that previous theoretical models have neglected such a term entirely. In principle, one could also introduce a separate reservoir accounting for electron-hole pairs that conserve OAM, however, we assume that the dynamics of such a population can be neglected given that $a$ is small. Fig.~\ref{Figure2}(\textbf{a}) and (\textbf{b}) show the typical intensity and phase obtained in a stationary state from propagating Eqs.~\ref{Equation1} from a random initial condition.

Repeating with different randomly generated initial conditions showed similar results, where in Fig.~\ref{Figure2}(\textbf{c}) we show the number of vortex states attained out of $50$ realizations. It can be seen that vortices appear deterministically when $a$ exceeds a small threshold around $10^{-4}$. The vortices are also reasonably robust against disorder in the system, which was modelled by introducing a randomly generated Gaussian correlated potential into Eq.~\ref{Equation1} (with correlation length $0.5\mu m$ and root mean squared amplitude $V$). As shown in Fig.~\ref{Figure2}(\textbf{d}), so long as the disorder is below some critical limit, the vortices remain. It is notable that the tolerated disorder exceeds typical values reported experimentally (e.g., $0.1meV$ in Ref.~\cite{Krizhanovskii2006}).
\begin{figure}[b]
\includegraphics[width=1\columnwidth]{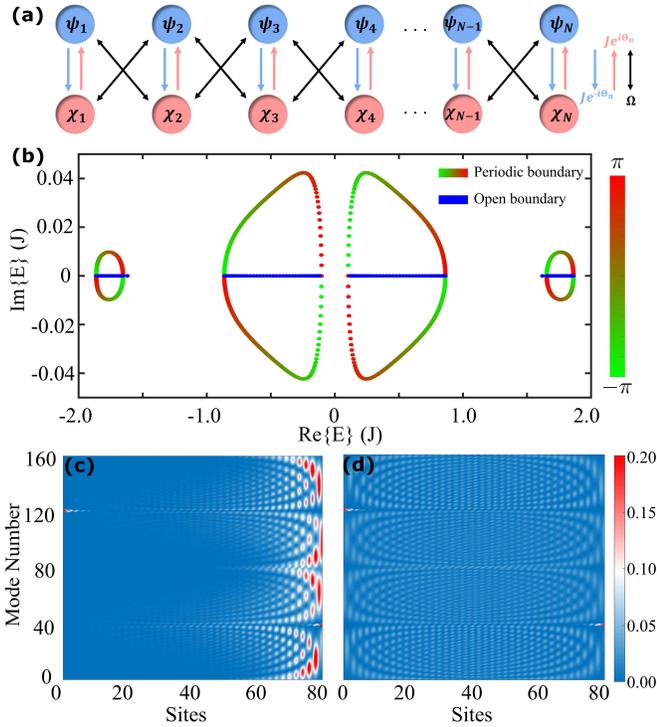}
\caption{\textbf{a}) The schematic for N coupled rings, where $\psi_n$/$\chi_n$ is the vortex/anti-vortex mode. The coupling strength between $\psi_n$ and $\chi_n$ in neighbouring rings is indicated by $\Omega$, while in the same ring the coupling from $\psi_n$ to $\chi_n$ is $Je^{-i\Theta_n}$ and $Je^{i\Theta_n}$ for $\chi_n$ to $\psi_n$. \textbf{b}) Eigenenergies calculated under periodic boundary and open boundary conditions. Parameters: $\Omega/J=0.5, \Delta/J=0.5,W/J=0.1,\gamma/J=0.2,\theta_{n=1,2,3,4}=(n-1)\pi/2$. \textbf{c}) Modes distribution under open boundary condition. Parameters: $\gamma=2W=0.2J$. \textbf{d}) Modes distribution under open boundary condition for $\gamma=2W=0$. For the finite case we have used 80 rings.}
\label{Figure3}
\end{figure}

{\it Unidirectional propagation in a lattice.} We now consider the behaviour of polaritons in a one-dimensional chain of coupled rings, where each ring is subjected to a non-resonant OAM carrying pump.
For an isolated ring, due to the rotational symmetry of the ring shaped potential, its eigenstates and eigenvalues appear in pairs (see SM Ref.~\cite{SupportingInformation}). There is a degeneracy of clockwise and anti-clockwise modes, which can also superimpose to form petal-shaped symmetric or antisymmetric combinations, which are also degenerate. To break the degeneracy of these states, we introduce an optical Gaussian defect in the ring potential and consequently couple the clockwise and anticlockwise modes (a similar coupling has been considered in Ref.~\cite{Xue2021}). We will consider first a simplified coupled mode analysis with a tight binding model for $N$ rings, where a vortex state ($\psi$) and an anti-vortex state ($\chi$) exist in each ring (see in Fig.~\ref{Figure3}(\textbf{a})). Later, we will consider the full spatial dynamics using the GP equation. The coupled mode equations are:
\begin{align}
i\hbar \frac{\partial \psi_{n}}{\partial t}=& ((-1)^{n+1}\Delta+iW) \psi_{n} + Je^{-i\Theta_{n}}\chi_{n}+\Omega(\chi_{n-1}+\chi_{n+1})  \notag \\ 
 i\hbar \frac{\partial \chi_{n}}{\partial t}=& ((-1)^{n+1}\Delta - \frac{i\gamma}{2})\chi_{n} + Je^{i\Theta_{n}}\psi_{n}+\Omega(\psi_{n-1}+\psi_{n+1}), 
\label{Equation2}
\end{align}
where $\psi_n$, $\chi_n$ are the wavefunctions associated to the clockwise and anti-clockwise modes, respectively, in the ring $n$,  $J$ is the coupling strength between the clockwise and anti-clockwise modes in the same ring, $\Omega$ is the coupling between the clockwise and anti-clockwise modes in neighbouring rings. $\Delta$ defines an energy detuning between rings, where we take an alternating detuning throughout the lattice. We will show later in the continuous model that this can be defined by choosing the size of the considered Gaussian defects. $\Theta_n$ is the angular position of the defect in the ring $n$. Note that the dependence on the angular position of the defect can be derived by assuming that it splits the symmetric and antisymmetric combinations of clockwise and anticlockwise modes in energy (see SM Ref.~\cite{SupportingInformation}). We assume that both clockwise and anti-clockwise modes experience a loss $\gamma$, but only clockwise modes experience a pumping $P$ making their net gain amplitude $W=(P-\gamma)/2$. We consider the limit where $|W|\ll \gamma/\Omega$, such that the anti-clockwise mode ($\chi_n$) has a fast dynamics compared to the clockwise modes ($\psi_n$). This limit is not essential and will be relaxed later when we consider full spatial dynamics, but for the purpose of illustration it is helpful in allowing to adiabatically eliminate the $\chi_n$ modes by approximating them as stationary. Note that Eq.~\ref{Equation2} should be supplemented by open boundary conditions $\psi_0=\chi_0=\psi_{N+1}=\chi_{N+1}=0$ when finite chains are considered.

Setting $d\chi_n/dt =0$ and substituting the solutions for $\chi_n$ into the equations of $\psi_n$, one may notice that $\psi_1$ to $\psi_N$ are coupled and we can write the following effective coupling equation $i\hbar \frac{\partial}{\partial t}
\Psi =\mathcal{H}\Psi$, where $\Psi = [\psi_1, \psi_2,...,\psi_N]^{T}$ and the elements in $\mathcal{H}$ can be written as:
\begin{align}
&\mathcal{H}_{n,n}=iW+\Delta(-1)^{n+1}-\frac{J^2}{(-1)^{n+1}\Delta-i\gamma/2}-\frac{2\Omega^2}{(-1)^{n}\Delta-i\gamma/2} , \notag \\
&\mathcal{H}_{n+1,n}=\frac{-J\Omega e^{-i\Theta_n} }{(-1)^{n+1}\Delta-i\gamma/2} + \frac{-J\Omega e^{i\Theta_{n+1}} }{(-1)^{n}\Delta-i\gamma/2}, \notag \\
&\mathcal{H}_{n,n+1}=\frac{-J\Omega e^{i\Theta_n} }{(-1)^{n+1}\Delta-i\gamma/2} + \frac{-J\Omega e^{-i\Theta_{n+1}} }{(-1)^{n}\Delta-i\gamma/2}, \notag \\
&\mathcal{H}_{n,n+2}=\frac{-\Omega^2}{-\Delta-i\gamma/2},\mathcal{H}_{n+2,n}=\frac{-\Omega^2}{\Delta-i\gamma/2}.
\label{Equation3}
\end{align}
Details of obtaining Eqs.~\ref{Equation3} are shown in SM Ref.~\cite{SupportingInformation}. In the case of open boundary conditions, $\mathcal{H}_{1,1}$ and $\mathcal{H}_{N,N}$ take a slightly different form with $2\Omega^2$ replaced with $\Omega^2$. The Hamiltonian (3) already shows an emergent non-Hermitian behaviour in our system, where $\mathcal{H}_{n+1,n}\neq\mathcal{H}_{n,n+1}^*$. 

By properly choosing the positions of the defects ($\Theta_n$), we can set the coupling $H(n+1,n)$ to zero while keeping $H(n,n+1)$ non-zero and therefore achieve the non-reciprocal coupling between the modes in separate rings. Following this principle, we get the condition that $e^{i(\Theta_{n}+\Theta_{n+1})} = (\Delta\pm i\gamma/2)/(\Delta\mp i\gamma/2)$ for odd and even $n$ respectively. To show that our system is topological we define a topological invariant, known as the winding number in Brillouin zone, following~\cite{Gong2018,Ghatak2019,Bergholtz2021,Borgnia2020}:
\begin{align}
\nu=\sum_{l=1}^{N_l}\int_{-\pi}^{\pi}\frac{dk}{2\pi}\partial_k \arg [E_l(k)-E_P],
\label{Equation4}
\end{align}
where $\arg[E_l(k)]$ is the argument of the complex energy $E_l(k)$ calculated from the non-Hermitian Bloch Hamiltonian $H(k)$ and $N_l$ corresponds to the total number of bands (see SM Ref.~\cite{SupportingInformation} for details). The complex energy spectrum does not cross the reference point $E_P$ i.e., $H(k)$ is point gapped with respect to $E_P$~\cite{Gong2018}. The non-trivial winding number corresponds to the total number of times the complex energy encircles the point $E_P$. For $\nu \neq 0$ the system is topologically non-trivial with the consequence that the modes are localized at one edge of a finite system. $\nu=\pm 1$ also corresponds to loops when the real energy of $H(k)$ is plotted against the imaginary energy (see Fig.~\ref{Figure3}(\textbf{b}))). Since the unit cell forms with four rings we get  four loops in the complex energy spectrum.  For $\nu= 1$ the loop winds in the counter clockwise direction with the modes being localized at the left end of the finite system, while $\nu= -1$ corresponds to a clock-wise winding with the modes being localized at the right end. In our case, all the loops have clockwise winding ($\nu= -1$) and the modes are localized at the right end of the finite lattice (see Fig.~\ref{Figure3}(\textbf{c})). The spectrum of the periodic boundary is also drastically different from the open boundary, which is a signature of the breakdown of bulk-boundary correspondence associated with skin modes~\cite{Yang2020,Okuma2020,Zhang2020,Yao2018,Zhu2020} (see Fig.~\ref{Figure3}(\textbf{b})).  Since all the modes are localized at the right end of the lattice, polaritons always propagate towards the right, while propagation towards the left is strongly supressed. However, such non-reciprocal behaviour  vanishes if we remove the non-Hermitian pump and decay term from the system ($\gamma=W=0$). In this case, the spatial profile of the modes changes drastically, where the modes become delocalized into the bulk and the system becomes reciprocal (see Fig.~\ref{Figure3}(\textbf{d})).

Note that in contrast to previous works operating with the non-Hermitian skin effect, we do not require non-Hermitian or non-reciprocal coupling between lattice sites directly~\cite{Li2020,Weidemann2020,Helbig2020,Zhang2021}. All the coupling between modes in our system is Hermitian, while the local gain and loss at different lattice sites makes the system non-Hermitian. Effectively, the interplay of these local non-Hermitian components with the phase-dependent Hermitian coupling makes an effective non-Hermitian coupling between the considered lattice sites. 

To show that our scheme does not require the previously used approximations (tight-binding, neglect of reservoir dynamics, and adiabatic elimination of $\chi_n$ modes), we simulate the unidirectional propagation in a one dimensional ring array (consisting of five polariton rings) in the continuous model, where each ring is excited by an OAM carrying non-resonant pump.  Note that in the tight-binding model, we accounted for a pair of clockwise and anticlockwise states in each ring, while in the continuous model there is a ladder of clockwise and anticlockwise states corresponding to different values of the angular momentum $l$. We choose parameters so as to arrange the desired effect for a particular value of $l=4$. In principle, the scattering to other angular momenta could occur due to the presence of disorder in the system, however, we find that in practice such scattering is negligible for typical disorder potential characteristics. The initial signal is introduced as a short non-resonant pulse on the first/last ring. To verify the validity of the approximations made in the tight-binding model, we write the GP equations corresponding to the system as
\begin{align}
i\hbar \frac{\partial \psi(\bm{r},t)}{\partial t}= (-\frac{\hbar^2 \nabla^2}{2m}+ g |\psi(\bm{r},t)|^2 +g_R n_R(\bm{r},t)  \notag \\ +V(r) + \frac{i\hbar}{2}(Rn_R(\bm{r},t)-\gamma))\psi(\bm{r},t) 
\notag \\+i(F_0+f)\iint \phi^{*}(\bm{r'}) \psi(\bm{r'},t)d\bm{r}' \phi(\bm{r}), \notag \\
\frac{\partial n_R(\bm{r},t)}{\partial t}= (-\gamma_R + R|\psi(\bm{r},t)|^2)n_R + P(\bm{r})+f,
\label{Equation6}
\end{align}
where $V(r)$ is the potential profile with disorder, $F_0$ is the amplitude of the OAM preserving non-resonant pump and $f$ is a pulse added on the first ring; other terms are the same as in Eq.~\ref{Equation1}.

\begin{figure}[h!]
\includegraphics[width=1\columnwidth]{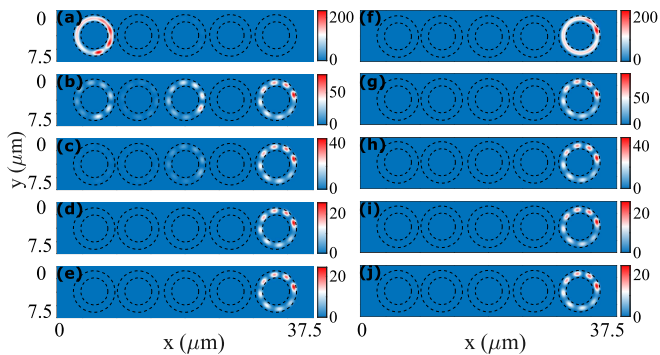}
\caption{\textbf{a}) Intensity distribution ($\mu m^{-2}$) of the wavefunction $\psi$ at $t=20$ ps, when the pulse is added at the left first ring. \textbf{b}), \textbf{c}), \textbf{d}), and \textbf{e}) represent $|\psi|^2$ at $t=$ 60, 100, 140, 180ps respectively. Disorder parameters: root mean square amplitude: 0.1$meV$; correlation length: 0.5 $\mu m$. \textbf{f}) to \textbf{j}) Corresponding intensity distribution when the pulse is introduced on the right first ring. Alternative detuning throughout the lattices is done by placing defects with alternating strength in neighbouring rings. This suggest that propagation towards right is allowed while towards left is suppressed, demonstrating the unidirectional propagation through the lattice.}
\label{Figure4}
\end{figure}

Figs.~\ref{Figure4}(\textbf{a}) to (\textbf{e}) show the results when the pulse is added on the left first ring. As seen, the off-resonant pulse (input signal) is introduced on the left ring first at $t=20ps$, then the intensity travels forward without backscattering even in the presence of disorder. When the evolution time reaches more than $300$ $ps$ (Fig.~\ref{Figure4}(\textbf{d})), the system is steady and the signal stays at the last ring. The ratio of the intensity in the first and last ring over the total intensity in the whole space is also calculated, which indicates that $100\%$ of the signal is being transmitted (see SM Ref.~\cite{SupportingInformation}). Also, when the pulse is added on the last ring, as can be seen in Fig.~\ref{Figure4}(\textbf{e}) to (\textbf{h}), the whole intensity of the signal stays on the last ring and no signal is read on the first one. This result in the continuous model is consistent with the interpretation attained earlier with the tight-binding model. Note that our scheme is not limited to the lifetime of polaritons. The presence of gain in the system allows signals to propagate far further than the distance that a single polariton can travel ballistically before decaying. Although we considered five rings in the demonstration of Fig.~\ref{Figure4}, the same non-reciprocal propagation can be arranged for any number of rings.

\emph{Conclusion.}---A recent experiment~\cite{Kwon2019} has shown that the orbital angular momentum (OAM) of a non-resonant laser can be conserved during the formation of an exciton-polariton condensate. This motivates a correction to the generalized mean-field theory of polariton condensation. We introduce such a correction, in the form of OAM conserving driving, which we find only needs to be relatively small to match the experimental observations. We further predict that the OAM driving term is sufficient to induce a non-reciprocal one-way transport in a one-dimensional lattice of polariton rings, provided that defects are engineered at specific locations in the rings to introduce a phase dependent coupling between clockwise and anticlockwise modes in each ring. This effect can be considered as a consequence of the vanishing of specific matrix elements in an effective tight-binding model or equivalently as a consequence of a topological skin effect due to the non-Hermitian nature of the system. We anticipate that such a mechanism can be useful in the growing development of schemes for computation based on polariton angular momentum modes~\cite{Sigurdsson2014,Ma2017,Gao2018,Ma2020}, where a one-way feedback free coupling of basis states has not yet been available, In future work it would be interesting to consider the interplay of polariton lattices with two-site pumping and dissipation schemes that have been proposed in general quantum systems with structured reservoirs~\cite{Keck2018}. Although in the present work we have restricted ourselves to one dimensional topological phases, our system is suitable for realizing two dimensional topological phases~\cite{Zhang2021,Liang2013,Gao2016}.

\vspace{0.5cm}
\emph{Acknowledgement.}
This work was supported by the Singaporean Ministry of Education (MOE) via the Tier 2 Academic Research Fund project MOE2019-T2-1-004.

\bibliography{references}
\end{document}